**Enhancing Spin Coherence in Optically Addressable Molecular Qubits through Host-Matrix Control**

S. L. Bayliss[1,2†], P. Deb[1,3†], D. W. Laorenza[4,5†], M. Onizhuk[6], G. Galli[1,6,7], D. E. Freedman[4*], D. D. Awschalom[1,3,7*]

[1]Pritzker School of Molecular Engineering, University of Chicago, Chicago, IL 60637, USA
[2]James Watt School of Engineering, University of Glasgow, Glasgow G12 8QQ, United Kingdom
[3]Department of Physics, University of Chicago, Chicago, IL 60637, USA
[4]Department of Chemistry, Massachusetts Institute of Technology, Cambridge, Massachusetts 02139, USA
[5]Department of Chemistry, Northwestern University, Evanston, IL 60208, USA
[6]Department of Chemistry, University of Chicago, Chicago, IL 60637, USA
[7]Center for Molecular Engineering and Materials Science Division, Argonne National Laboratory, Lemont, IL 60439, USA

[†]These authors contributed equally to this work
*awsch@uchicago.edu, danna@mit.edu

**Abstract**

Optically addressable spins are a promising platform for quantum information science due to their combination of a long-lived qubit with a spin-optical interface for external qubit control and read out. The ability to chemically synthesize such systems—to generate optically addressable *molecular* spins—offers a modular qubit architecture which can be transported across different environments, and atomistically tailored for targeted applications through bottom-up design and synthesis. Here we demonstrate how the spin coherence in such optically addressable molecular qubits can be controlled through engineering their host environment. By inserting chromium (IV)-based molecular qubits into a non-isostructural host matrix, we generate noise-insensitive clock transitions, through a transverse zero-field splitting, that are not present when using an isostructural host. This host-matrix engineering leads to spin-coherence times of more than 10 μs for optically addressable molecular spin qubits in a nuclear and electron-spin rich environment. We model the dependence of spin coherence on transverse zero-field splitting from first principles and experimentally verify the theoretical predictions with four distinct molecular systems. Finally, we explore how to further enhance optical-spin interfaces in molecular qubits by investigating the key parameters of optical linewidth and spin-lattice relaxation time. Our results demonstrate the ability to test qubit structure-function relationships through a tunable molecular platform, and highlight opportunities for using molecular qubits for nanoscale quantum sensing in noisy environments.

**I. Introduction**

Quantum information science offers potential to revolutionize computational, sensing, and networking capabilities [1–5]. Innovations in these areas depend, in part, on optimization of the quantum bit, or qubit, the fundamental unit of quantum information processing, and its environment. Solid-state color centers (such as the nitrogen-vacancy center in diamond) offer a well-developed qubit platform with a robust spin-optical interface for single qubit initialization and readout [3,6–8], and microwave-frequency spin transitions for coherent control. Introducing these properties into tunable and portable molecular systems combines the key properties of solid-state color centers with opportunities for optimization through bottom-up engineering of both the qubit and its environment [9,10]. These optically addressable molecular spin systems, or 'molecular color centers', provide Angstrom-scale precision and tunability of the local qubit environment [11–15], offering targeted design for applications such as nanoscale quantum sensing [16–18]. Additionally, the ground state spin hosted in a molecule comprises a portable qubit of <1 nm$^3$ size, such that these systems can be integrated into various host matrices and hybrid materials architectures [19–21].

Altering the symmetry or strain environment of solid-state color centers has been shown to significantly enhance qubit properties [22–25]. The portability of molecular qubits opens up a versatile platform to achieve such enhancements by modifying a qubit's electronic structure through host-matrix tuning, without altering the chemical composition of the qubit itself. In particular, the sensitivity of the ground-state spin to its local environment suggests that the crystallographic symmetry of the host matrix can be used to induce 'clock transitions' which are first-order insensitive to magnetic-field noise, and hence enhance spin coherence, even in noisy— e.g., nuclear spin-rich—environments. As demonstrated in molecular qubits without a spin-optical interface [26–28] and solid-state spin systems [29–31], such transitions can enhance spin coherence, reducing the need for isotopic control of the nuclear spin environment (e.g., deuteration [32]) or high qubit dilution to prolong coherence.

Here, we create clock transitions in optically addressable molecular spin qubits through host-matrix induced symmetry control. We use the chromium-based molecular color center—Cr(IV)(*o*-tolyl)$_4$, **1-Cr** [Fig. 1(a)]—to demonstrate the impact of the host environment on the ground state spin structure. The crystallographic symmetry of **1-Cr** and its isostructural, diamagnetic host, Sn(IV)(*o*-tolyl)$_4$, **1-Sn**, yields ground-state spin transitions that are first-order sensitive to magnetic-field fluctuations [13]. In contrast, and as we outline below, inserting **1-Cr** into a non-isostructural, lower symmetry host matrix— Sn(IV)(4-fluoro-2-methylphenyl)$_4$, **2-Sn** [Fig. 1(b)]— induces clock transitions as a result of the creation of a significant transverse zero-field splitting. These host-induced clock transitions significantly enhance the spin coherence of **1-Cr** in **2-Sn** (which we refer to as **2**) compared to **1-Cr** in the isostructural **1-Sn** host (which we refer to as **1**). We model this behavior from first principles using generalized cluster-correlation expansion methods, and further experimentally demonstrate enhanced optical contrast and spin-lattice relaxation times for these host-matrix engineered molecular color centers. Remarkably, the host modification to achieve this coherence enhancement comprises interchange of just one hydrogen atom on the host ligands with a fluorine atom. Thus, the coherence enhancement arises purely from symmetry control by the host—without requiring control of the nuclear spin bath—offering a pathway for coherence-protected quantum sensing (e.g., of electric fields and strain) in intrinsically noisy environments (e.g., biological systems), all within a versatile molecular platform.

## II. Results and discussion

### A. Host-guest control of molecular color centers

We recently demonstrated optical addressability of molecular spin qubits comprising a chromium (Cr$^{4+}$) ion coordinated by organic ligands in a pseudo-tetrahedral geometry such as **1-Cr** in Fig. 1(a) [13]. The symmetry and tetravalent oxidation state of **1-Cr** leads to a spin-triplet ($S$=1) ground state while the strong field organic ligands generate a suitable energy level structure for optical spin initialization and readout. As a result, the ground state can be optically initialized and read out using spin-selective excitation to the spin-singlet ($S$=0) excited state, combined with photoluminescence (PL) detection, analogous to solid-state color centers [see Fig. 1(c)] [33,34]. Specifically, resonantly exciting these molecular spin qubits with a narrow-linewidth laser initializes the ground state through optical pumping: selective excitation of a spin sublevel (e.g., $|0\rangle$ in Fig. 1(c), combined with excited-state decay, transfers population to the other spin sublevels [e.g., $|\pm\rangle$ in Fig. 1(c)]. Similarly, the same selective excitation enables optical spin readout through the resulting PL: spin sublevels resonant with the laser will be excited and give rise to PL, while the detuned spin sublevels will be only weakly excited and therefore give rise to weak (ideally vanishing) PL. Combined with microwave control of the ground state spin, these properties therefore provide a qubit which can initialized, coherently controlled, and read out in a fashion which is compatible with single-spin detection.

For **1**, the axial zero-field splitting is $D$=3.63 GHz, which splits the $m=0$ sublevel, $|0\rangle$, from the $m=\pm 1$ sublevels, $|\pm 1\rangle$, Fig. 1(b). In the tetragonally symmetric crystal environment of **1**, the Cr$^{4+}$ site contains a four-fold improper rotational axis ($S_4$) that enforces the transverse zero-field splitting, $E$, to be $\simeq$0, resulting in spin transitions that are first-order sensitive to magnetic fields [Fig. 1(b)]. However, if $|E|$>0, the degeneracy of the zero-field states is broken: the $|\pm 1\rangle$ states hybridize to form the non-degenerate levels $|\pm\rangle = \frac{1}{\sqrt{2}}(|+1\rangle \pm |-1\rangle)$, which have no first-order magnetic moment. Therefore, systems with $|E|$>0 exhibit transitions which are first-order insensitive to magnetic fields around $B$=0 [Fig. 1(b)], where $B$ is an external magnetic field. To achieve $|E|$>0 with **1-Cr**, we introduce **1-Cr** into a non-isostructural host matrix, **2-Sn** [Fig. 1(b)], that contains no three-fold (or higher) rotational axes such that $E$ is no longer symmetry-constrained to zero. Here, we investigate the influence of this broken symmetry on the resulting spin structure and coherence times of this system.

**1-Cr** was synthesized as described previously [35] and diluted in either **1-Sn** [36] or **2-Sn** to form dilute single crystals of **1** and **2** respectively. Individual crystals of **1** and **2** were mounted on a microwave coplanar waveguide inside an optical cryostat at ~4 K (see Supplemental Material for further details [37]). We first determine the optical structure of **2** through PL measurements: upon off-resonant excitation (785 nm), we measure PL from the spin-singlet excited state to the spin-triplet ground state [Fig. 1(d)]. Similar to **1**, the spectrum of **2** shows a zero-phonon line (ZPL) at 1016 nm (compared to 1025 nm for **1** [13]), and a resolved phonon side band. Fig. 1(d) shows the photoluminescence excitation (PLE) spectrum taken by sweeping a narrow-linewidth laser across this ZPL, and detecting photons from the phonon side band, from which we extract an inhomogeneous broadening of $\simeq$50 GHz (full-width half-maximum). For all following experiments, we address a subensemble of molecules from this inhomogeneous distribution by exciting at the ZPL maximum with a narrow-line laser, and detect the emitted photons from the phonon side band.

### B. Host-matrix induced clock transitions in a molecular color center

We then determine the ground state zero-field splitting parameters of **2** through continuous-wave optically detected magnetic resonance (cw-ODMR). Under continuous optical excitation, applying a microwave frequency on resonance with a transition between spin sublevels of **2** increases the

PL due to the mixing of the 'bright' and 'dark' spin sublevels, i.e., the sublevels which are resonant with and detuned from the laser respectively. Fig. 2(a) shows the cw-ODMR spectrum as a function of both magnetic field and microwave frequency, from which we extract $D$=5.55 GHz, $E$=1.85 GHz. Importantly, compared to **1**—in which $E$=0—the use of a lower symmetry host matrix in **2** generates a significant transverse zero-field splitting, despite the minor modification—i.e., single-site fluorination—to the host-matrix ligands. In fact, **2** displays the largest possible transverse zero-field splitting for its axial zero-field parameter i.e., $|E|=|D|/3$, highlighting the significant symmetry breaking afforded by the host matrix. We note that in this system where $|E|=|D|/3$, two of the spin transitions are degenerate at zero field, we observe two resonances—at $D-E=2E$ and $D+E$. This degeneracy is lifted under an applied magnetic field. We further note that since the sign of $D$ and $E$ are not determined in our experiments, and do not influence our results, we take $D, E>0$ for concreteness. The field-frequency ODMR map [Fig. 2(a)] highlights the insensitivity of the spin transitions to magnetic field: to first order the energy of the $|0\rangle$, $|-\rangle$, and $|+\rangle$ sublevels do not shift with increasing field, in contrast to the linear Zeeman shift exhibited for $E$=0 in **1**.

## C. Host-matrix enhanced spin coherence

We now illustrate how this behavior significantly enhances the spin coherence ($T_2$) in **2** compared to **1**. As a pre-requisite for optically detected spin coherence measurements, we performed pulsed ODMR by applying the pulse sequence outlined in Fig. 2(c). A laser pulse initializes the subensemble of spins, which are then rotated by a microwave $\pi$-pulse before being read out through the PL with a second laser pulse. Fig. 2(c) shows the pulsed ODMR spectrum as a function of microwave frequency at $B$=0 for the (doubly degenerate) low-frequency transitions (at $D-E=2E$=3.7 GHz). We find an ODMR contrast of 40% (defined such that the maximum possible contrast is 100%, see Supplemental Material [37]), which is approximately an order of magnitude improvement from our previous demonstration of pulsed ODMR with **1** . To measure the spin coherence of **2**, we replace the microwave $\pi$-pulse in the pulsed ODMR sequence with a Hahn-echo—i.e., $\frac{\pi}{2} - \tau - \pi - \tau$ sequence, where $2\tau$ is the free-evolution time—followed by an additional $\pi/2$ pulse to project coherences onto populations for optical readout [Fig. 3(b)]. At zero magnetic field, we measure a ground state spin coherence time $T_2$=10.6±0.2 µs, despite the nuclear spin rich environment, and the relatively high Cr concentration (~1%, $\simeq$1-2x10$^{19}$ Cr$^{4+}$/cm$^3$). In contrast, we measure a significantly shorter $T_2$=2.0±0.1 µs for **1** at zero-magnetic field, indicating the effectiveness of the clock transition in **2** for enhancing spin coherence.

To further understand the dependence of zero-field spin coherence on the transverse zero-field splitting, we investigated two other molecular systems: Cr(IV)(2,3-dimethylphenyl)$_4$ diluted in Sn(2,3-dimethylphenyl)$_4$ (**3**), and Cr(IV)(2,4-dimethylphenyl)$_4$ diluted in Sn(2,4-dimethylphenyl)$_4$ (**4**). The additional methyl group on the ligands of these compounds induces lower symmetry crystal packing than **1**, and consequently, $E\simeq$0.5 GHz in both cases [1], providing additional testbeds of the role of the transverse zero-field splitting in enhancing spin coherence, here from tuning the qubit rather than the host matrix. Fig. 3(c) plots the zero-field coherence time for **1**, **2**, and **3** showing that $T_2$ increases with increasing $E$. **4** shows a similar $T_2$ as **3** due to its very similar $E$ value (see Supplemental Material [37]). Generally, these four systems highlight paths to engineer even longer coherence times through independently optimizing both the host matrix and the chemical composition of the qubit.

To further understand the spin coherence of these molecular color centers interacting with their nuclear spin bath we used first principles generalized cluster correlation expansion (gCCE) calculations with Monte Carlo bath state sampling using the PyCCE package [38]. Starting from the crystal structure for these compounds, we calculated the electron-nuclear hyperfine couplings of the Cr-containing molecule using density functional theory (DFT). Using DFT-computed spin densities, we calculate the interactions between the Cr center and nuclear spins in the host matrix and use point dipole-dipole interactions between nuclear spins [37]. The calculated zero-field $T_2$ as a function of transverse zero-field splitting $E$ shows good agreement with the experimental values [Fig. 3(c)]. Since the calculations only consider the nuclear spin bath, they highlight that Cr electronic spins are not a major limitation on the coherence in our experiments. Interestingly, the calculations also allow us to determine the distance at which nuclear spins play a significant role in determining the coherence. By varying the number of nearest-neighbor molecules included in the calculations, we found $T_2$ converges when 3-4 nearest neighbors are included, corresponding to a radius of approximately 1.5 nm around the Cr center (see Supplemental Material for further details [37]).

To the best of our knowledge, the behavior of ground-state electronic spin coherence in the low magnetic field regime (from 0 to ~100 mT) has largely been unexplored in molecular systems, but, as demonstrated in solid-state color centers, is an important domain for applications in quantum information science [39,40]. To explore this regime in molecular color centers, we measured $T_2$ as a function of magnetic field for **1**, **2**, and **3**. In each case, $T_2$ decreases with increasing magnetic field between 0 and 30 mT [Fig. 3(d)]. The good agreement between the experimental data and the gCCE calculations indicates this behavior arises from the magnetic-field dependent dynamics of the nuclear spin bath, combined with moving away from the clock transitions of **2** and **3**. A similar drop in $T_2$ with field due to nuclear spin bath dynamics—albeit at a lower field scale—has been studied in the nitrogen-vacancy center in diamond (which has $E$=0) [41]. In this case, $T_2$ drops with magnetic field to a minimum at ~0.1 mT due to nuclear spin bath dynamics, before recovering at ~10 mT, when the nuclear Zeeman splitting dominates over the electron-nuclear and nuclear-nuclear spin interactions [41]. Our observations in molecular color centers can be assigned to a similar mechanism—albeit with a larger characteristic field scale due to the stronger spin interactions—combined with the increased noise sensitivity for **2** and **3** as they shift from their zero-field clock transitions. Due to the higher field scale involved for these molecular color centers, we only see a reduction in $T_2$ in the measured field range. We note, however, that electron spin resonance measurements performed on **1** at higher magnetic fields (≃400 mT) yield $T_2$≃2.5 μs [15], consistent with the theoretical prediction of a high-field recovery in $T_2$ [41] (See Supplemental Material for further details [37]).

### D. Key parameters to optimize molecular spin-optical interfaces

Having shown how host-matrix engineering can enhance the coherence of molecular color centers, we now explore additional key properties of **2**. The optical linewidth is a crucial parameter in these molecular systems: it determines the readout and initialization fidelity by setting the spin selectivity of the excitation. Quantifying this linewidth is therefore an important step to further optimize molecular color centers. To measure the homogeneous optical linewidth (i.e., that of the subensemble of spins probed under resonant excitation), we perform a two laser tone experiment. We apply a fixed laser tone, at frequency $f_L$, along with a second laser tone, detuned by $\Delta f_L$, which we sweep. When the difference in laser frequencies, $\Delta f_L$, matches the spin transition frequencies, the second laser tone excites population that was shelved in other (dark) spin sublevels, thus increasing the PL. Similarly, when $\Delta f_L=0$, the PL is lower than for a finite detuning since the population has already been shelved in the dark sublevels by the first laser tone. The linewidths of these spectral peaks and holes enable us to determine the homogeneous optical linewidth of the subensemble—which determines the spin-optical contrast—from the inhomogeneously broadened ensemble (Fig. 4). To mitigate the slope in the PL traces [Fig. 4(c)] caused by the inhomogeneous broadening, we use a differential ODMR measurement: we apply a fixed microwave drive $f_{MW}$ at the frequency of one of the spin transitions, and measure the ODMR signal as we sweep $\Delta f_L$. From a fit to these measurements, we extract an optical linewidth of $\simeq 3$ GHz for **2** [Fig. 4(d) - see Supplemental Material for further details [37]]. Importantly, since this is comparable to the zero-field splitting parameters, this indicates promise for significantly improving molecular spin-optical interfaces by lowering linewidths. Future work will focus on understanding the electron-phonon and electron-electron dephasing mechanisms contributing to this linewidth (*i.e.,* optical coherence) [42]. We note that similar spin-flip, intraconfigurational optical transitions in coordination compounds have exhibited homogeneous linewidths on the order of 10-100 MHz [43,44], suggesting avenues for future improvements through e.g., lower temperatures, selective deuteration [45], or reduced $Cr^{4+}$ concentration.

We next measure the resulting optical contrast of **2**—which provides a lower bound on the spin polarization— by applying an optical pulse (2 ms long) and measuring the emitted photons during this pulse, followed by a wait time much greater than the spin-lattice relaxation time for ground state equilibration before the next repetition of the experiment. Over the course of the optical pulse, the PL decreases as spins are optically pumped from the probed 'bright' spin sublevel to the 'dark' spin sublevels. The difference in PL at the beginning and the end of the pulse provides a lower bound on the spin polarization of 65% [Fig. 5(a)]. This is a marked improvement on **1**, where we previously observed a contrast of 14% [13]. We then measure the spin-lattice relaxation time $T_1$ of the ground-state spin by applying an optical initialization pulse, followed by a variable relaxation time, $T$, before measuring the relaxation of the spins through a read-out pulse during which we collect the PL. By varying the relaxation time, we measure $T_1=1.21\pm0.02$ ms [Fig. 5(b)], a five-fold improvement compared to our previous measurements of **1**.

The enhanced optical contrast and $T_1$ of **2** suggests either improved thermalization and/or crystal quality relative to **1**. For example, measuring $T_1$ as a function of temperature for **2** shows that $T_1$ drops by more than a factor of four between 4 K and 7 K (see Supplemental Material [37]), highlighting the dramatic influence of temperature on $T_1$. Thus, a five-fold enhancement of $T_1$ for **2** relative to **1** could result from a decrease in effective temperature at the sample of few Kelvin. Similarly, a reduction in local temperature should reduce the optical linewidths [46], thus improving the contrast. Additionally, the increased $|D|$ of **2** can improve the spin selectivity of resonant optical excitation. Thus, improving sample thermalization and modifying $|D|$ to larger, yet still measurable, values offer clear routes to optical contrast approaching 100% for molecular color centers. Overall, these measurements show how key molecular qubit properties can be engineered through host-matrix control.

## III. Conclusion

This work demonstrates how symmetry engineering through atomistic host-matrix control of a molecular spin qubit can significantly enhance both spin coherence and spin-optical interfaces. Through host-based symmetry control, we have demonstrated spin coherence times exceeding 10 μs for optically addressable molecular spins in a nuclear and electron spin-rich environment. These results combine the advantages of noise-protected coherence with optical qubit initialization, coherent control, and readout in a versatile molecular architecture. By exploring the influence of the host-matrix on optical contrast, homogeneous optical linewidths, and spin-lattice relaxation times, our results highlight further directions to improve molecular spin-optical interfaces e.g., by using lower temperatures to limit thermal vibrations, and engineering vibrational modes through isotope control and modification of the chemical makeup of the host matrix or qubit. The ability to transport molecular qubits between different environments, and tune these hosts with atomic level precision, highlights exciting opportunities for further control over a range of qubit properties e.g., spin-orbit coupling [47], as well as integration with photonic or phononic devices [48,49]. In particular, the enhanced coherence demonstrated at zero magnetic field could be used to sense electric fields, strain, or temperature at the nanoscale, while retaining insensitivity to magnetic-field noise.

Our results demonstrate the promise of rapidly advancing molecular color centers through an iterative feedback loop of bottom-up design, targeted chemical synthesis, qubit measurement, and accurate first-principles calculations. For example, our results point to even longer coherence using $E$ values of ~10 GHz, which should be synthetically achievable [50]. Such $E$ values would enable accessible ground-state spin control, while further enhancing optical contrast through increased ground-state spin splittings. Overall, the flexibility of molecular color center design combined with the application of accurate theoretical tools offer promise for optimizing spin-optical interfaces, for applications ranging from nanoscale quantum sensing to long-distance entanglement distribution.

**Acknowledgements.** *Funding.* We acknowledge funding from ONR N00014-17-1-3026 and the MRSEC Shared User Facilities at the University of Chicago (NSF DMR-1420709). D.W.L. and D.E.F thank US Army Research Office under award number W911NF2010088 for the synthesis. D.E.F. and D.D.A. acknowledge support for the integration of molecules into a larger quantum infrastructure from the U.S. Department of Energy Office of Science National Quantum Information Science Research Centers. *Competing interests.* S.L.B., D.W.L., D.E.F., and D.D.A. are inventors on patent application no. 63008589 submitted by the University of Chicago that covers chemically tunable optically addressable molecular-spin qubits and associated methods. S.L.B., P. D., D.W.L., M.O., G.G., D.E.F., and D.D.A. are inventors on patent application no. 22-T-121 by the University of Chicago that covers enhancing quantum properties of optically addressable molecular qubits through host matrix engineering. *Data and materials availability.* The data can be accessed at [URL to be added in proof]. Crystallographic data can be obtained free of charge via www.ccdc.cam.ac.uk/data_request/cif or by emailing data_request@ccdc.cam.ac.uk. CCDC codes: [to be added in proof].

**References**

[1] C. L. Degen, F. Reinhard, and P. Cappellaro, *Quantum Sensing*, Rev. Mod. Phys. **89**, 1 (2017).

[2] S. Wehner, D. Elkouss, and R. Hanson, *Quantum Internet: A Vision for the Road Ahead*, Science **362**, (2018).

[3] D. D. Awschalom, R. Hanson, J. Wrachtrup, and B. B. Zhou, *Quantum Technologies with Optically Interfaced Solid-State Spins*, Nat. Photonics **12**, 516 (2018).

[4] H.-S. Zhong, H. Wang, Y.-H. Deng, M.-C. Chen, L.-C. Peng, Y.-H. Luo, J. Qin, D. Wu, X. Ding, Y. Hu, P. Hu, X.-Y. Yang, W.-J. Zhang, H. Li, Y. Li, X. Jiang, L. Gan, G. Yang, L. You, Z. Wang, L. Li, N.-L. Liu, C.-Y. Lu, and J.-W. Pan, *Quantum Computational Advantage Using Photons*, Science **370**, 1460 (2020).

[5] D. Awschalom, K. K. Berggren, H. Bernien, S. Bhave, L. D. Carr, P. Davids, S. E. Economou, D. Englund, A. Faraon, M. Fejer, S. Guha, M. V. Gustafsson, E. Hu, L. Jiang, J. Kim, B. Korzh, P. Kumar, P. G. Kwiat, M. Lončar, M. D. Lukin, D. A. B. Miller, C. Monroe, S. W. Nam, P. Narang, J. S. Orcutt, M. G. Raymer, A. H. Safavi-Naeini, M. Spiropulu, K. Srinivasan, S. Sun, J. Vučković, E. Waks, R. Walsworth, A. M. Weiner, and Z. Zhang, *Development of Quantum Interconnects (QuICs) for Next-Generation Information Technologies*, PRX Quantum **2**, 1 (2021).

[6] M. Atatüre, D. Englund, N. Vamivakas, S. Y. Lee, and J. Wrachtrup, *Material Platforms for Spin-Based Photonic Quantum Technologies*, Nat. Rev. Mater. **3**, 38 (2018).

[7] G. Wolfowicz, F. J. Heremans, C. P. Anderson, S. Kanai, H. Seo, A. Gali, G. Galli, and D. D. Awschalom, *Quantum Guidelines for Solid-State Spin Defects*, Nat. Rev. Mater. **6**, 906 (2021).

[8] M. Ruf, N. H. Wan, H. Choi, D. Englund, and R. Hanson, *Quantum Networks Based on Color Centers in Diamond*, J. Appl. Phys. **130**, 070901 (2021).

[9] M. Atzori and R. Sessoli, *The Second Quantum Revolution: Role and Challenges of Molecular Chemistry*, J. Am. Chem. Soc. **141**, 11339 (2019).

[10] M. R. Wasielewski, M. D. E. Forbes, N. L. Frank, K. Kowalski, G. D. Scholes, J. Yuen-Zhou, M. A. Baldo, D. E. Freedman, R. H. Goldsmith, T. Goodson, M. L. Kirk, J. K. McCusker, J. P. Ogilvie, D. A. Shultz, S. Stoll, and K. B. Whaley, *Exploiting Chemistry and Molecular Systems for Quantum Information Science*, Nat. Rev. Chem. **4**, 490 (2020).

[11] M. S. Fataftah and D. E. Freedman, *Progress towards Creating Optically Addressable Molecular Qubits*, Chem. Commun. **54**, 13773 (2018).

[12] M. S. Fataftah, S. L. Bayliss, D. W. Laorenza, X. Wang, B. T. Phelan, C. B. Wilson, P. J. Mintun, B. D. Kovos, M. R. Wasielewski, S. Han, M. S. Sherwin, D. D. Awschalom, and D. E. Freedman, *Trigonal Bipyramidal $V^{3+}$ Complex as an Optically Addressable Molecular Qubit Candidate*, J. Am. Chem. Soc. **142**, 20400 (2020).

[13] S. L. Bayliss, D. W. Laorenza, P. J. Mintun, B. Diler, D. E. Freedman, and D. D. Awschalom, *Optically Addressable Molecular Spins for Quantum Information Processing*, Science **370**, 1309 (2020).

[14] M. K. Wojnar, D. W. Laorenza, R. D. Schaller, and D. E. Freedman, *Nickel(II) Metal Complexes as Optically Addressable Qubit Candidates*, J. Am. Chem. Soc. **142**, 14826 (2020).

[15] D. W. Laorenza, A. Kairalapova, S. L. Bayliss, T. Goldzak, S. M. Greene, L. R. Weiss, P. Deb, P. J. Mintun, K. A. Collins, D. D. Awschalom, T. C. Berkelbach, and D. E. Freedman,

*Tunable Cr$^{4+}$ Molecular Color Centers*, J. Am. Chem. Soc. **143**, 21350 (2021).

[16] G. Kucsko, P. C. Maurer, N. Y. Yao, M. Kubo, H. J. Noh, P. K. Lo, H. Park, and M. D. Lukin, *Nanometre-Scale Thermometry in a Living Cell*, Nature **500**, 54 (2013).

[17] R. Schirhagl, K. Chang, M. Loretz, and C. L. Degen, *Nitrogen-Vacancy Centers in Diamond: Nanoscale Sensors for Physics and Biology*, Annu. Rev. Phys. Chem. **65**, 83 (2014).

[18] J. F. Barry, M. J. Turner, J. M. Schloss, D. R. Glenn, Y. Song, M. D. Lukin, H. Park, and R. L. Walsworth, *Optical Magnetic Detection of Single-Neuron Action Potentials Using Quantum Defects in Diamond*, Proc. Natl. Acad. Sci. U. S. A. **113**, 14133 (2016).

[19] S. Thiele, F. Balestro, R. Ballou, S. Klyatskaya, M. Ruben, and W. Wernsdorfer, *Electrically Driven Nuclear Spin Resonance in Single-Molecule Magnets*, Science **344**, 1135 (2014).

[20] G. Czap, P. J. Wagner, F. Xue, L. Gu, J. Li, J. Yao, R. Wu, and W. Ho, *Probing and Imaging Spin Interactions with a Magnetic Single-Molecule Sensor*, Science **364**, 670 (2019).

[21] I. Cimatti, L. Bondì, G. Serrano, L. Malavolti, B. Cortigiani, E. Velez-Fort, D. Betto, A. Ouerghi, N. B. Brookes, S. Loth, M. Mannini, F. Totti, and R. Sessoli, *Vanadyl Phthalocyanines on Graphene/SiC(0001): Toward a Hybrid Architecture for Molecular Spin Qubits*, Nanoscale Horizons **4**, 1202 (2019).

[22] A. Barfuss, J. Teissier, E. Neu, A. Nunnenkamp, and P. Maletinsky, *Strong Mechanical Driving of a Single Electron Spin*, Nat. Phys. **11**, 820 (2015).

[23] E. R. Macquarrie, T. A. Gosavi, S. A. Bhave, and G. D. Fuchs, *Continuous Dynamical Decoupling of a Single Diamond Nitrogen-Vacancy Center Spin with a Mechanical Resonator*, Phys. Rev. B - Condens. Matter Mater. Phys. **92**, 224419 (2015).

[24] Y. I. Sohn, S. Meesala, B. Pingault, H. A. Atikian, J. Holzgrafe, M. Gündoğan, C. Stavrakas, M. J. Stanley, A. Sipahigil, J. Choi, M. Zhang, J. L. Pacheco, J. Abraham, E. Bielejec, M. D. Lukin, M. Atatüre, and M. Lončar, *Controlling the Coherence of a Diamond Spin Qubit through Its Strain Environment*, Nat. Commun. **9**, 2012 (2018).

[25] K. C. Miao, J. P. Blanton, C. P. Anderson, A. Bourassa, A. L. Crook, G. Wolfowicz, H. Abe, T. Ohshima, and D. D. Awschalom, *Universal Coherence Protection in a Solid-State Spin Qubit*, Science **369**, 1493 (2020).

[26] M. Shiddiq, D. Komijani, Y. Duan, A. Gaita-Ariño, E. Coronado, and S. Hill, *Enhancing Coherence in Molecular Spin Qubits via Atomic Clock Transitions*, Nature **531**, 348 (2016).

[27] J. M. Zadrozny, A. T. Gallagher, T. D. Harris, and D. E. Freedman, *A Porous Array of Clock Qubits*, J. Am. Chem. Soc. **139**, 7089 (2017).

[28] K. Kundu, J. R. K. White, S. A. Moehring, J. M. Yu, J. W. Ziller, F. Furche, W. J. Evans, and S. Hill, *Clock Transition Due to a Record 1240 G Hyperfine Interaction in a Lu(II) Molecular Spin Qubit*, ChemRxiv (2021).

[29] G. Wolfowicz, A. M. Tyryshkin, R. E. George, H. Riemann, N. V. Abrosimov, P. Becker, H. J. Pohl, M. L. W. Thewalt, S. A. Lyon, and J. J. L. Morton, *Atomic Clock Transitions in Silicon-Based Spin Qubits*, Nat. Nanotechnol. **8**, 561 (2013).

[30] A. Ortu, A. Tiranov, S. Welinski, F. Fröwis, N. Gisin, A. Ferrier, P. Goldner, and M. Afzelius, *Simultaneous Coherence Enhancement of Optical and Microwave Transitions in Solid-State Electronic Spins*, Nat. Mater. **17**, 671 (2018).

[31] M. Onizhuk, K. C. Miao, J. P. Blanton, H. Ma, C. P. Anderson, A. Bourassa, D. D.

Awschalom, and G. Galli, *Probing the Coherence of Solid-State Qubits at Avoided Crossings*, PRX Quantum **2**, 010311 (2021).

[32] C. J. Wedge, G. A. Timco, E. T. Spielberg, R. E. George, F. Tuna, S. Rigby, E. J. L. McInnes, R. E. P. Winpenny, S. J. Blundell, and A. Ardavan, *Chemical Engineering of Molecular Qubits*, Phys. Rev. Lett. **108**, 107204 (2012).

[33] W. F. Koehl, B. Diler, S. J. Whiteley, A. Bourassa, N. T. Son, E. Janzén, and D. D. Awschalom, *Resonant Optical Spectroscopy and Coherent Control of $Cr^{4+}$ Spin Ensembles in SiC and GaN*, Phys. Rev. B **95**, 035207 (2017).

[34] B. Diler, S. J. Whiteley, C. P. Anderson, G. Wolfowicz, M. E. Wesson, E. S. Bielejec, F. Joseph Heremans, and D. D. Awschalom, *Coherent Control and High-Fidelity Readout of Chromium Ions in Commercial Silicon Carbide*, Npj Quantum Inf. **6**, 11 (2020).

[35] S. U. Koschmieder, B. S. McGilligan, G. McDermott, J. Arnold, G. Wilkinson, B. Hussain-Bates, and M. B. Hursthouse, *Aryl and Aryne Complexes of Chromium, Molybdenum, and Tungsten. X-Ray Crystal Structures of [Cr(2-MeC6H4)(μ-2-MeC6H4)(PMe3)]2, Mo(H2-2-MeC6H3)(2-MeC6H4)2(PMe2Ph)2, and W(H2-2,5-Me2C6H2)(2,5-Me2C6H3)2-(PMe3)2*, J. Chem. Soc. Dalt. Trans. 3427 (1990).

[36] C. Schneider-Koglin, B. Mathiasch, and M. Dräger, *Über Tetraaryl-Methan-Analoga in Der Gruppe 14. III. $Ar_4Sn/Pb$ (Ar Ph, p-, m-, o-Tol, 2,4-Xyl Und 2,5-Xyl): Gegenüberstellung von Bindungslängen Und Winkeln, von NMR Chemischen Verschiebungen Und Kopplungskonstanten Und von Schwingungsdaten*, J. Organomet. Chem. **469**, 25 (1994).

[37] *See Supplemental Material at [URL to Be Added in Proof] for Further Details*.

[38] M. Onizhuk and G. Galli, *PyCCE: A Python Package for Cluster Correlation Expansion Simulations of Spin Qubit Dynamics*, Adv. Theory Simulations **4**, 2100254 (2021).

[39] H. Bernien, B. Hensen, W. Pfaff, G. Koolstra, M. S. Blok, L. Robledo, T. H. Taminiau, M. Markham, D. J. Twitchen, L. Childress, and R. Hanson, *Heralded Entanglement between Solid-State Qubits Separated by Three Metres*, Nature **497**, 86 (2013).

[40] F. Shi, Q. Zhang, P. Wang, H. Sun, J. Wang, X. Rong, M. Chen, C. Ju, F. Reinhard, H. Chen, J. Wrachtrup, J. Wang, and J. Du, *Single-Protein Spin Resonance Spectroscopy under Ambient Conditions*, Science **347**, 1135 (2015).

[41] N. Zhao, S. W. Ho, and R. B. Liu, *Decoherence and Dynamical Decoupling Control of Nitrogen Vacancy Center Electron Spins in Nuclear Spin Baths*, Phys. Rev. B - Condens. Matter Mater. Phys. **85**, 115303 (2012).

[42] H. Riesen, *Hole-Burning Spectroscopy of Coordination Compounds*, Coord. Chem. Rev. **250**, 1737 (2006).

[43] H. Riesen and E. Krausz, *Persistent Spectral Hole-Burning, Luminescence Line Narrowing and Selective Excitation Spectroscopy of the R Lines of Cr(III) Tris(2,2'-Bipyridine) in Amorphous Hosts*, J. Chem. Phys. **97**, 7902 (1992).

[44] J. L. Hughes and H. Riesen, *Zeeman Effects in Transient Spectral Hole-Burning of the R1 Line of $NaMgAl(Oxalate)_3 \cdot 9H_2O$/Cr(III) in Low Magnetic Fields*, J. Phys. Chem. A **107**, 35 (2003).

[45] C. Wang, S. Otto, M. Dorn, E. Kreidt, J. Lebon, L. Sršan, P. Di Martino-Fumo, M. Gerhards, U. Resch-Genger, M. Seitz, and K. Heinze, *Deuterated Molecular Ruby with Record Luminescence Quantum Yield*, Angew. Chemie - Int. Ed. **57**, 1112 (2018).

[46] D. Serrano, K. S. Kumar, B. Heinrich, O. Fuhr, D. Hunger, M. Ruben, and P. Goldner, *Rare-

*Earth Molecular Crystals with Ultra-Narrow Optical Linewidths for Photonic Quantum Technologies*, ArXiv (2021).

[47] J. Liu, J. Mrozek, A. Ullah, Y. Duan, J. J. Baldoví, E. Coronado, A. Gaita-Ariño, and A. Ardavan, *Quantum Coherent Spin–Electric Control in a Molecular Nanomagnet at Clock Transitions*, Nat. Phys. **17**, 1205 (2021).

[48] Y. Tian, P. Navarro, and M. Orrit, *Single Molecule as a Local Acoustic Detector for Mechanical Oscillators*, Phys. Rev. Lett. **113**, 135505 (2014).

[49] C. Toninelli, I. Gerhardt, A. S. Clark, A. Reserbat-Plantey, S. Götzinger, Z. Ristanović, M. Colautti, P. Lombardi, K. D. Major, I. Deperasińska, W. H. Pernice, F. H. L. Koppens, B. Kozankiewicz, A. Gourdon, V. Sandoghdar, and M. Orrit, *Single Organic Molecules for Photonic Quantum Technologies*, Nat. Mater. **20**, 1615 (2021).

[50] M. Rubín-Osanz, F. Lambert, F. Shao, E. Rivière, R. Guillot, N. Suaud, N. Guihéry, D. Zueco, A. L. Barra, T. Mallah, and F. Luis, *Chemical Tuning of Spin Clock Transitions in Molecular Monomers Based on Nuclear Spin-Free Ni(II)*, Chem. Sci. **12**, 5123 (2021).

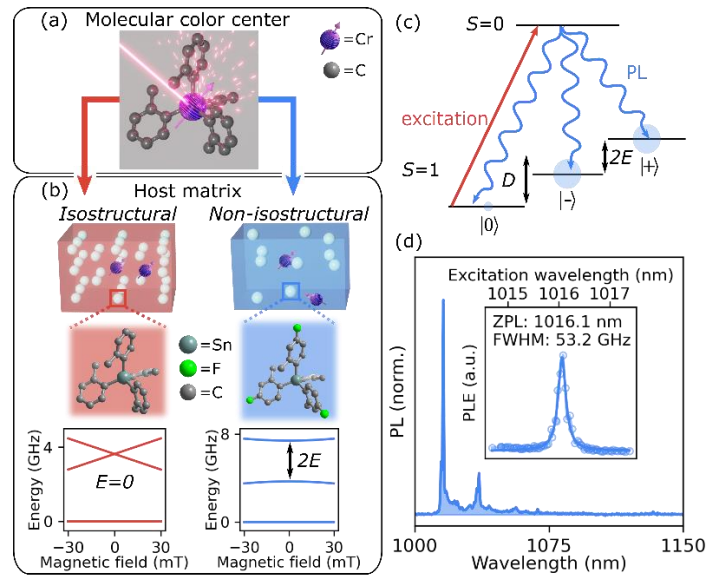

Figure 1. Host-matrix engineering of optically addressable molecular qubits. (a) Molecular structure of **1-Cr** (determined from single crystal X-ray diffraction) with laser excitation and emission outlined. Hydrogen atoms are omitted for clarity. (b) Single crystal packing diagram of **1-Cr** in its isostructural host, **1-Sn** (red, left), and non-isostructural host, **2-Sn** (blue, right), showing positions of metal centers. Hydrogen atoms are omitted for clarity. The resulting ground-state spin structures show the clock transition ($E>0$) induced in **2**. (c) Energy level diagram of chromium molecular color centers, highlighting resonant excitation to, and photoluminescence (PL) from, the $S=0$ excited state, and zero-field splitting of the ground-state spin sublevels. (d) PL and photoluminescence excitation (PLE) spectra of **2** at 4 K.

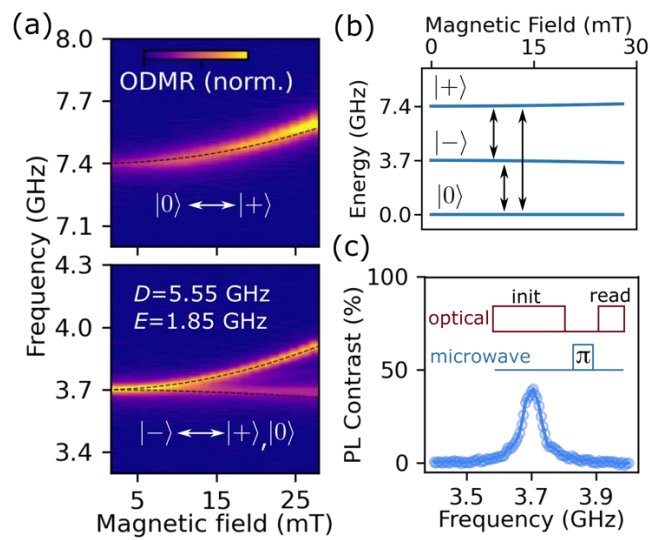

Figure 2. Host-matrix engineered clock transitions in **2**. (a) Continuous-wave optically detected magnetic resonance (cw-ODMR) as a function of magnetic field and microwave frequency overlaid with simulated spin transition frequencies (dashed lines). The ODMR spectrum yields $D$=5.55 GHz, $E$=1.85 GHz. (b) Calculated spin-sublevel energies as a function of magnetic field highlighting their magnetic-field insensitivity. (c) Pulsed ODMR at zero magnetic field demonstrating an optical contrast of approximately 40%. Inset: pulsed ODMR sequence comprising optical initialization (init), microwave, and optical read out (read) pulses.

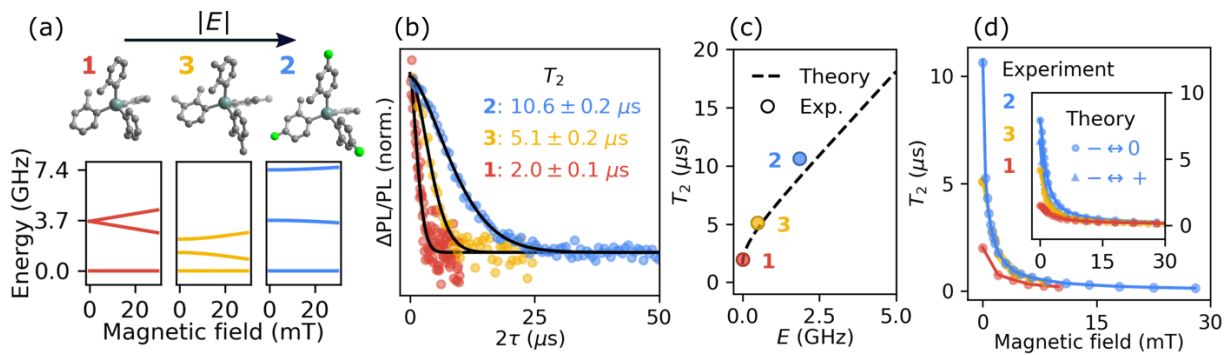

Figure 3. Host and chemical tuning of transverse zero-field splitting to enhance coherence. (a) Molecular structures of the host matrix for **1**, **2**, and **3** with their simulated spin energy levels as a function of magnetic field. (b) Hahn echo traces for **1**, **2**, and **3** at zero magnetic field. (c) Zero-field spin coherence as a function of transverse zero-field splitting along with theoretical values. (d) Experimental and calculated $T_2$ as a function of magnetic field.

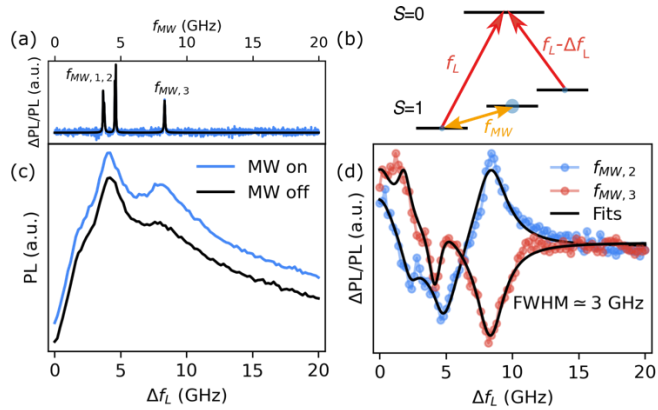

Figure 4. Extracting the homogeneous optical linewidth for **2**. (a) ODMR showing the relevant microwave transitions ($B\sim70$ mT). (b) Schematic of the two-color experiments showing a fixed laser tone, $f_L$, along with a sideband detuned by $\Delta f_L$, which is swept (d) PL as a function of $\Delta f_L$ with and without a microwave drive ($f_{MW,2} \simeq 4.6$ GHz). (e) ODMR as a function of $\Delta f_L$ for applied microwave tones at $f_{MW,2}$ and $f_{MW,3}$ ($\simeq 8.3$ GHz) with fits yielding a homogeneous optical linewidth of $\simeq 3$ GHz.

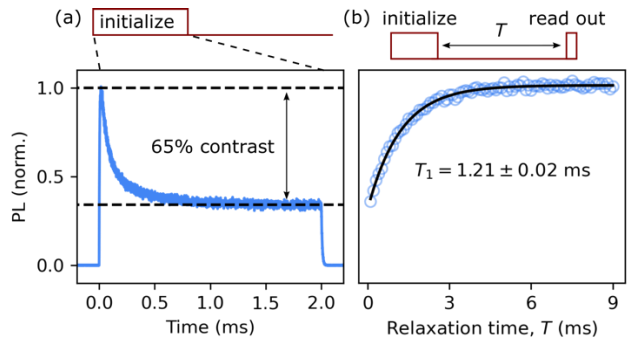

Figure 5. Enhanced optical contrast and spin-lattice relaxation time for host-matrix engineered molecular qubits. (a) Optical spin initialization observed as a reduction in PL over the course of an applied laser pulse. (b) Spin-lattice relaxation time measured by an all-optical sequence comprising initialization and readout laser pulses, separated by a variable relaxation time $T$.